# Reentrant Superconductivity in Strontium Titanate Micropillars Under Elastic Stress

**Jacob Franklin[1,♦], Joshua Bedard[1,♦], Brenna Petrelli[1] and Ilya Sochnikov[1,2,3,♣]**

[1]*Department of Physics, University of Connecticut, Storrs, CT, USA 06269*

[2]*Institute of Material Science, University of Connecticut, Storrs, CT, USA 06269*

[3]*Department of Material Science & Engineering, University of Connecticut, Storrs, CT, USA 06269*

♦*Equal contributors*

♣*Corresponding author; e-mail: ilya.sochnikov@uconn.edu*

Abstract

**A superconducting pairing mechanism in carrier-doped $SrTiO_3$ remains a fundamental open question. Carefully tuning the material towards its quantum ferroelectric critical point is essential for studying this unconventional superconductivity because of the strong connection between the superconducting and quantum critical ferroelectric states. Strain is a promising tuning parameter, but past experiments on strain-tuned quantum-critical $SrTiO_3$ have yielded unexpected and inconsistent results. We address these challenges by performing scanning SQUID experiments on compressed micropillars of superconducting $SrTiO_3$. Our results reveal two distinct behaviors of the AC susceptibility under compressive stress, wherein the susceptibility is first suppressed at low stress, but then enhanced at high stress. This reentrant superconductivity suggests that the micropillar experiences a domain reorientation and explains previous contradictory experiments in the context of strain-tuning within the quantum critical models of superconductivity. In addition, our newly developed technique provides a unique set of advantages for future experiments on stress- or strain-tuned quantum materials.**

The exact mechanism for the superconductivity in strontium titanate while doped with carriers remains a highly active area of research due to the rich physics that emerges from the interplay between its quantum states [1–21]. Superconductivity in doped $SrTiO_3$ exhibits several unique characteristics, such as its low carrier concentration and Fermi energy [22]. This low Fermi energy violates the assumption in BCS theory that the Debye energy must be much less than the Fermi energy [23]. In fact, doped $SrTiO_3$ has the largest ratio of Debye to Fermi energies among any superconductor [16]. In addition, $SrTiO_3$ exhibits an emergence of a superconducting dome as a function of carrier concentration [24], which is reminiscent of unconventional cuprate superconductors. While these characteristics clearly distinguish this material from BCS superconductors, the majority of modern theoretical models accept that the electron pairing is phonon-mediated [16]. These models have proposed a variety of explanations to reconcile phonon-mediated pairing with the contradictions to BCS theory, which include contributions from longitudinal modes [6], plasmon modes [3,5], dual phonon interactions [19,25], and spin-orbit coupling [26,27]. Quantum critical ferroelectric soft mode models [4,11,8,18–21] propose that quantum critical ferroelectric fluctuations [28–35] provide the pairing mechanism via the ferroelectric transverse optical (TO) phonon mode, and have been propelled to the forefront of attention due to recent experimental [17,36,37] and theoretical [18–21] works.

A major challenge that quantum critical models of superconductivity have faced is matching results from compressive strain experiments to their strict prediction that tuning away from the ferroelectric quantum critical point (QCP) will suppress the superconducting state [4]. This QCP is the zero kelvin point that divides the normal paraelectric phase and the ferroelectric phase which is suppressed by quantum fluctuations. Quantum paraelectric materials such as $SrTiO_3$ can be tuned toward or away from their ferroelectric QCP via strain. Stiffening the ferroelectric TO phonon mode via hydrostatic pressure tunes the material away from the ferroelectric QCP [38], and has consistently suppressed the superconducting critical temperature [39,40]. Experiments with *c*-axis compression should suppress the superconducting state for the same reasons, but the results are inconclusive: some experiments show an enhancement in the critical temperature [39,41], others a suppression [42], while yet other experiments have observed more complicated results that reveal both enhancement and suppression [37,39]. Conversely, tuning towards the QCP by softening the ferroelectric TO phonon mode via Sr substitution [43], isotope substitution of oxygen [15,44], and tensile strain [17,36] has consistently enhanced the superconducting critical temperature, as expected. An explanation is needed for why experimental results show that tuning towards the QCP via tensile strain enhances the superconducting state but do not consistently show that tuning away from the QCP via compressive strain suppresses the superconducting state.

In this manuscript, we address the open questions posed by previous $SrTiO_3$ compression experiments by performing scanning SQUID (Superconducting QUantum Interference Device) experiments on a compressively strained $SrTi_{0.996}Nb_{0.004}O_3$ pillar-shaped superelastic microstructure known as a micropillar [45]. This technique integrates highly tunable stress, the large homogenous elastic strains of microstructures, and quantum sensing. This combination of features has exciting applications for strain- and stress-tunable quantum materials. The stress-dependent measurements of the AC (alternating current) susceptibility in this superconducting micropillar exhibit an unexpected behavior which suggests an evolution in the pillar's structural domain distribution. Structural domain mechanics have scarcely been considered in past compression experiments on $SrTiO_3$, but as we show in this manuscript, they provide an explanation for why these experiments have produced inconsistent results. In light of the structural domain evolution suggested by the stress-tuned susceptibility measurements presented here, present and past results support quantum critical ferroelectric models of superconductivity in $SrTiO_3$.

In order to measure the continuous uniaxial stress dependence of the superconducting state in $SrTiO_3$, we integrated micropillars with our scanning SQUID microscope. First, a 5x5x0.5 $mm^3$ bulk sample of $SrTi_{0.996}Nb_{0.004}O_3$ was manually ground down to a thickness of ~30 μm using sandpaper and a crystal polisher. The carrier concentration of this sample was $6.4\times10^{19}$ $cm^{-3}$ [37]. Then, it was placed into a custom clamp so that it could be milled with a Helios xenon plasma focused ion beam (PFIB) dual beam system. Using a high beam current of 1.3 μA, we milled away about 35 μm from a large 150 μm long region at the corner of the sample except for a small location where the pillar would be formed. This location was further refined using progressive steps of lower beam currents (15 nA to 24 pA) until a vertical, smooth, cylindrical pillar remained with approximately 6 μm in diameter and 12 μm in height, see Fig. 1a. When milled with a PFIB, $SrTiO_3$ exhibits a 'terracing' effect [46] that poses an obstacle to achieving homogenous strain, which is why long sessions using very low beam currents where necessary to smooth the pillar. In addition, it was essential to have the pillar fabricated in a way that was largely isolated from nearby material so that it could be accessed by the pickup loop on our scanning SQUID. This is why there is such a large volume of the sample removed around the pillar, see Fig. 1c.

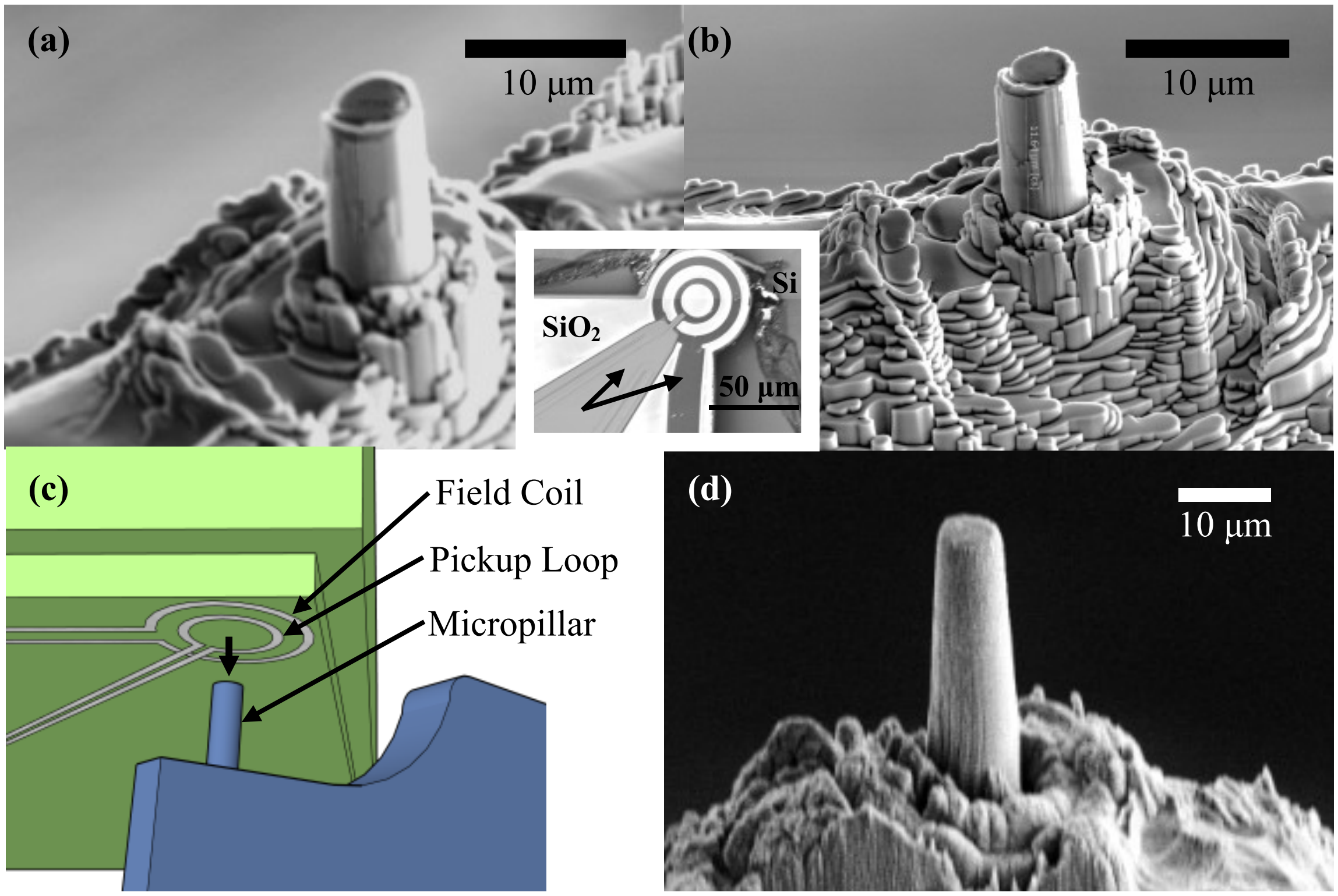

FIG. 1. Measurements were performed on a PFIB-milled $SrTi_{0.996}Nb_{0.004}O_3$ micropillar. (a) An SEM image of the pillar after the fabrication was complete. First, a large region was milled away from the corner of a thin bulk sample except for a small region near the tip. Then, this region was refined into a smooth, cylindrical pillar. The terrace-like features around the pillar result from the PFIB milling. These features were smoothed out from the sides of the pillar using very small beam currents as low as 24 pA. (b) An SEM image of the micropillar after the experiment was complete. After our scanning SQUID experiment compressed the pillar many times, the pillar shows no deformation. The inset between (a) and (b) shows an SEM image of the SQUID's pick-up loop and field coil after the experiment. The lack of any visible damage shows that the stress applied to the pillar was within the elastic limit of the SQUID's silicon chip. (c) A diagram of the SQUID over the pillar. In order to perform our stress-dependent susceptibility measurements, we located the pillar by scanning laterally over it. Then, we pressed down at the location of the pillar (meaning that the SQUID pickup loop is centered over the pillar) and measured the susceptibility as a function of the voltage on the piezoelectric scanner. The SQUID is mounted on a brass cantilever which measures the stress that the SQUID is exerting on the pillar. (d) A different micropillar made of titanium was fabricated and measured identically as a control sample.

After fabricating the $SrTi_{0.996}Nb_{0.004}O_3$ micropillar, we transferred the sample clamp to our scanning SQUID microscope, which consists of a SQUID magnetometer mounted on a custom-built scanner made from parallel arrays of piezoelectric bimorphs installed in a Bluefors LD250 dilution refrigerator [47]. A SQUID is a flux-to-voltage transducer that can achieve immense magnetic flux sensitivities of ~$10^{-7}\ \Phi_0/\sqrt{Hz}$ [48–51]. A SQUID also measures the diamagnetic response (AC susceptibility) using the mutual inductance between the field coil, which provides a local field source, and the pickup loop. The SQUID used in this experiment had a large pickup loop with a diameter of 20 μm. It was essential for our 6 μm diameter superconducting pillar to contact the SQUID entirely within the pickup loop so that electric shorts were not present. The pickup loop was on the surface of a bulk silicon chip, so when the pickup loop and pillar were concentric and contact was initiated, the top of the micropillar was in contact with a flat silicon surface that was coplanar with the pickup loop (see Fig. 1c). This prevented the conducting pillar from shorting the pickup loop.

To conduct AC susceptibility measurements on the micropillar under stress, we first positioned the SQUID with the pickup loop directly above the pillar using the lateral components of our piezoelectric scanner. The pickup loop was approximately 50 μm above the top of the pillar. Then we approached the pillar with the SQUID using the vertical component of the scanner and initiated contact between the top of the pillar and the interior of the SQUID pickup loop. When the SQUID came into contact with the pillar, it displaced a flexible brass cantilever on which it was mounted. This cantilever was connected to a capacitance bridge, which measures the capacitance between the cantilever and a nearby copper plate. This measures the displacement of the cantilever and was used to determine which piezoelectric voltage corresponds to the point of contact with the pillar, which was at approximately 4 V for all temperatures, see Fig. 2. In Fig. 2, this piezoelectric voltage is on the horizontal axis. Increasing the piezoelectric voltage corresponds to moving the SQUID towards the sample, which is downwards in the Fig. 1c. After contact, further voltage to the scanner induced stress onto the pillar. The capacitance signal (displacement of the flexible cantilever) was used to calculate the force exerted on the pillar, and thus the stress. The stress is nonlinear with voltage to the piezo and initially exhibits a step-like behavior, see Fig. S1 in Supplementary Information. We note that because the maximum stress achieved in this experiment is on the order of 0.1 GPa and the Young's modulus of strontium titanate is on the order of 100 GPa [52,53], the maximum strain in our pillar is on the order of 0.1%. During the approach and contact, the dilution refrigerator's pulse tube was temporarily turned off to reduce vibrations. The temperature of the sample stage remained stable for around 3 minutes after the pulse tube was turned off: enough time for one measurement. Throughout the entire process, the field coil provided an external AC magnetic field and the susceptibility of the pillar was measured by the SQUID pickup loop. This susceptibility signal represents the superconducting strength of the material comprising the pillar.

There are two distinct behaviors of the AC susceptibility as stress increases. The susceptibility signal is reduced (in amplitude) immediately after contact and continues decreasing for approximately 1 V, see Fig. 2, blue lines. This corresponds to a stress range of ~0 MPa to ~5 MPa, see Fig. 2, red lines. This decrease in the susceptibility amplitude indicates that the superconducting state in the pillar is being suppressed. For one temperature, 153 mK, the susceptibility amplitude was only reduced by 35%, but for all other temperatures a decrease of 70-100% was typical. For all temperatures, the susceptibility signal in this region is unstable. Further than ~1 V above the point of contact, the susceptibility amplitude suddenly changed its behavior and began sharply increasing. For every temperature it eventually exceeded its zero-stress value, which shows that the superconducting state in the pillar at high stress (~10 MPa to ~100 MPa) was enhanced compared to the state at zero stress. As stress increases further, susceptibility typically saturates, changing from a dramatically steep but somewhat irregular increase to a modest regular linear increase, which is maintained to the highest stresses we obtained. At our highest stresses of around 100 MPa, an enhancement of 400% compared to the susceptibility at 0 MPa was typical.

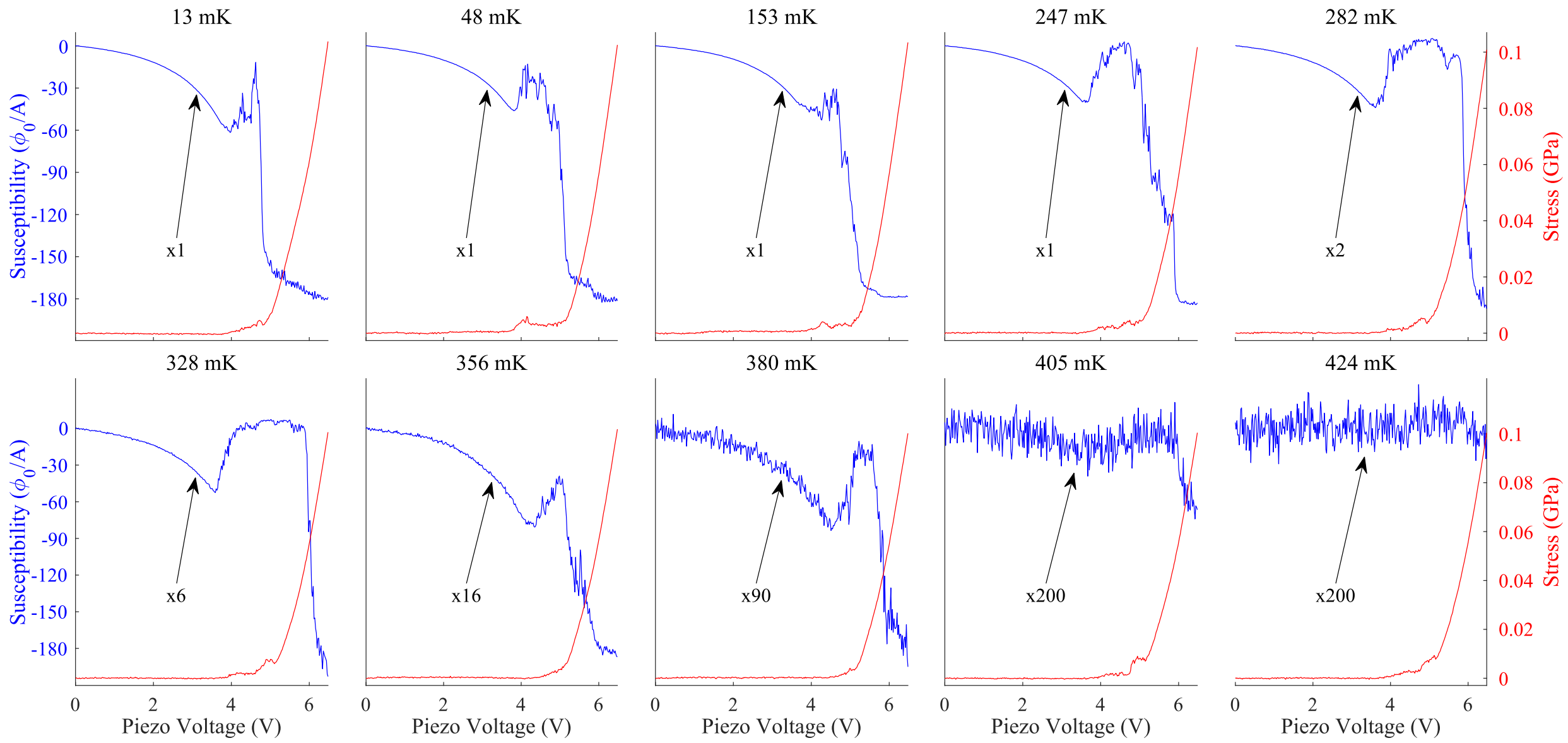

FIG. 2. Reentrant superconductivity in an $SrTiO_3$ micropillar: the superconducting response of the $SrTi_{0.996}Nb_{0.004}O_3$ micropillar is first suppressed and then enhanced under uniaxial compressive stress. The susceptibility of the pillar is measured as a function of voltage on the piezoelectric scanner, which induces stress on the pillar the above ~4 V. The susceptibility for each temperature is shown by the blue lines, whose scales appear to the left of each row. At higher temperatures, the susceptibility data is magnified for visual aid. The stress for each temperature is shown by the red lines, whose scales are on the right of each row. At voltages less than ~4 V, the SQUID is approaching the micropillar. As expected, susceptibility increases during the approach and stress remains zero. At ~4 V, the top of the pillar is in contact with the interior of the pickup loop. Above ~4 V, increasing the voltage on the piezoelectric scanner increases the stress on the pillar. For stress from 0 MPa to 100 MPa, two behaviors are apparent in the susceptibility data: at lower stress the susceptibility is suppressed and at higher stress the susceptibility is enhanced. Similarly, two behaviors are apparent in the stress data: at lower piezo voltages the stress increases marginally and is unstable but at higher piezo voltages the stress increases significantly and is stable. The nonmonotonic susceptibility and unstable stress are consistent with a domain reorientation and show that the impact of compression on the superconducting state depends on the orientation of the tetragonal *c*-axis.

The technique we developed, which integrates scanning probe microscopy with the unique properties of FIB-milled micropillars, is highly novel. To ensure that the complete suppression and enhanced reentrant superconducting behavior that we observe is not an experimental artifact, we performed the exact same procedure on a control sample, which was a micropillar made from pure (99.99%) titanium, see Fig. 1d. The Ti micropillar had a 6μm diameter like the Nb:$SrTiO_3$ micropillar. Our results on the Ti micropillar (Fig. 3) show only small effects of susceptibility due to contact with the SQUID and increasing pressure.

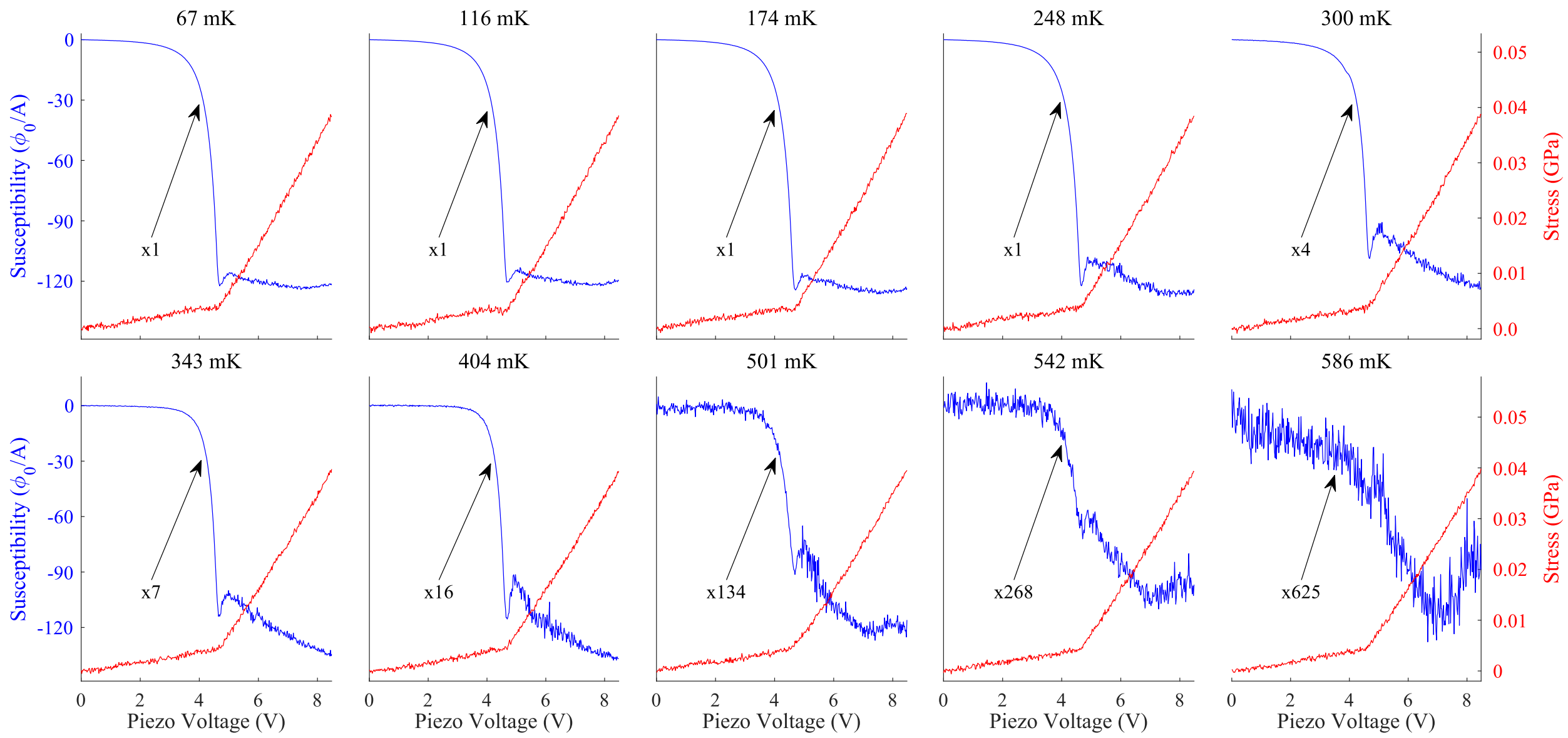

FIG. 3. Compression on a Ti micropillar reveals only small changes in susceptibility. Each subplot is a different temperature, with the lowest temperature on the top left and the highest temperature, above $T_c$, on the bottom right. In each plot, the red is the stress, whose scale is all the way to the right in each row. The blue is the susceptibility, with scales on the left of each row. The SQUID makes contact with the top of the pillar at ~4.5 V on the piezoelectric scanner. For each temperature, there is a small decrease in susceptibility amplitude upon contact with the SQUID, and then a slow increase back approximately the same susceptibility just before contact. The decrease of the amplitude is about ~5% of the total amplitude for lower temperatures. At higher temperatures, it can be as large as ~20%.

The reentrant behavior of the susceptibility of the $SrTi_{0.996}Nb_{0.004}O_3$ micropillar shows that not only is the superconducting strength (i.e., superfluid density) reduced at low stress and enhanced at high stress, but also the critical temperature is suppressed and enhanced in the same way. Under zero stress, we see a typical [24] superconducting critical temperature of ~380 mK. In Fig. 4a, this zero stress susceptibility for each temperature is the yellow line and was determined using the local maximum of the susceptibility amplitude near the point of contact in Fig. 2. Under a small amount of compressive stress (blue line in Fig. 4a), the critical temperature is suppressed to ~250 mK. These susceptibilities are determined by the local minimum of the susceptibility amplitude in Fig. 1 after the point of contact. Under high compressive stress (red line in Fig. 4a), the critical temperature is enhanced to >405 mK. The presence of both suppression and then enhancement in the superconducting state at different amounts of uniaxial compressive stress is unexpected if the macroscopic stress is applying homogenous strain. To explain these surprising observations, we interpret the reentrant superconductivity in the micropillar to be the result of an evolution in the structural domain distribution within the micropillar.

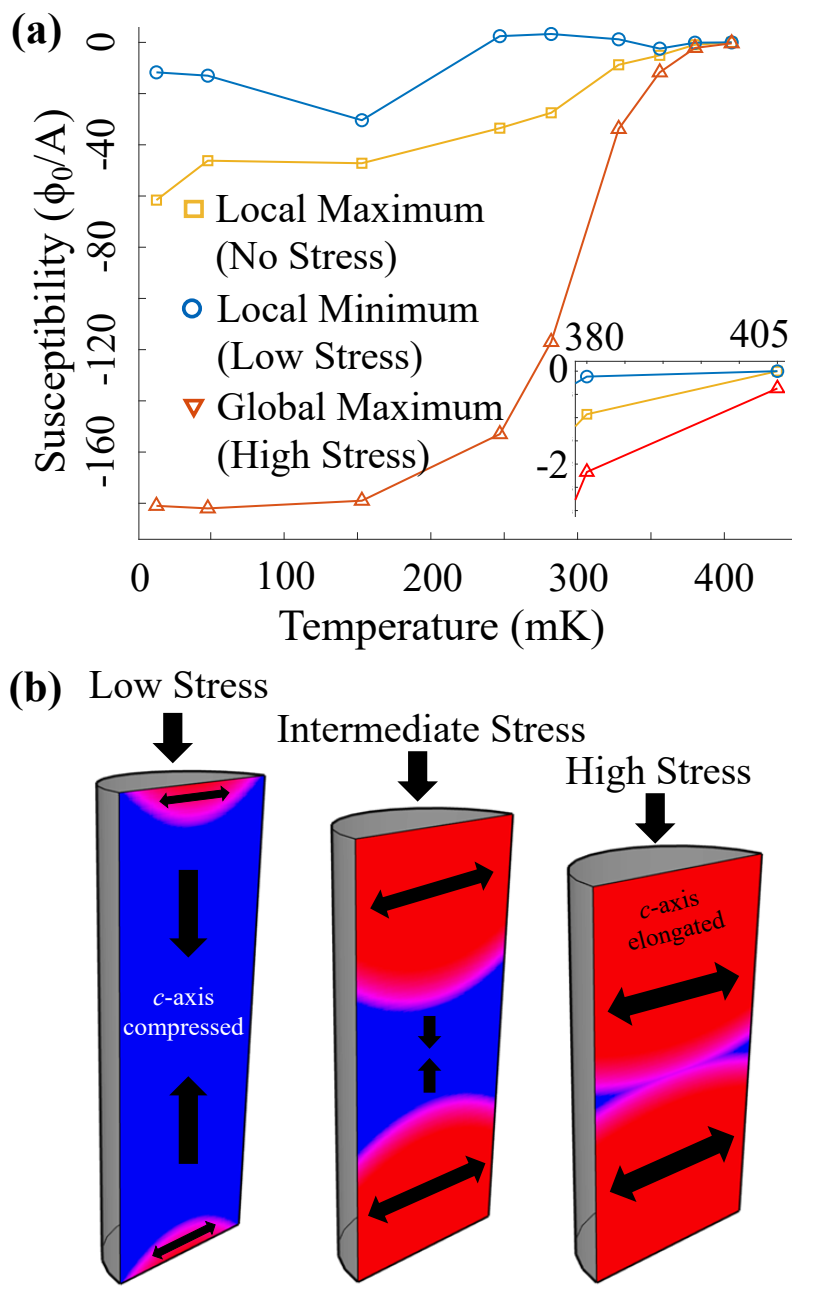

FIG. 4. Structural domains with differently oriented *c*-axes dictate the sign of strain on the *c*-axis and thus whether superconductivity is suppressed or enhanced under pillar compression. (a) The susceptibilities at key extrema from Fig. 2 are plotted here against temperature. At low stress (blue), the susceptibility is suppressed compared to zero stress (yellow). One temperature (153 mK) had a suppression of only 35%, but for all other temperatures a suppression of 80% was more typical, with some reaching 100%. At high stress (red), the susceptibility is enhanced by 300-400%. The inset shows a zoomed view near $T_c$, which shows that the $T_c$ is enhanced to >405 mK. (b) The suggested domain reorientation in the $SrTiO_3$ micropillar is illustrated here. At zero stress, micropillars of tetragonal crystals are dominated by domains with the *c*-axis aligned with the pillar axis (blue region). As compressive stress is applied, small regions near the ends of the pillar with c-axes aligned perpendicular to the pillar (red regions) spread and dominate. This domain reorientation would explain the unusual switch from suppressed superconductivity to enhanced superconductivity seen in (a). At low stresses, the superconductivity is suppressed because the *c*-axis is compressed. *c*-axis compression stiffens the ferroelectric TO phonon mode, which decreases the superconducting coupling constant and lowers the critical temperature. At high stresses, the domains have reoriented and the superconductivity is enhanced because the *c*-axis is stretched. *c*-axis elongation softens the ferroelectric TO phonon mode, which increases the coupling constant and increases the critical temperature.

It is well known that $SrTiO_3$ experiences a structural phase transition at 105 K from a higher temperature cubic phase to a lower temperature tetragonal phase. This phase transition is induced by an antiferrodistortive rotation of the oxygen octahedra in the perovskite unit cell, and results in an elongated *c*-axis with a c/a ratio of 1.00056 [54]. In single crystals, this tetragonal phase introduces [100], [010], and [001] structural domains corresponding to the elongated *c*-axis oriented along each of the three formerly cubic crystal directions [55]. These structural domains have been studied extensively because of the exotic effects that the small strains in domain walls have on the magnetism, ferroelectricity, conductivity, and superconductivity in $SrTiO_3$ bulk samples and heterostructures [56–63]. They are mobile even at low temperatures [59,64], and when $SrTiO_3$ is compressed with dozens of MPa they will align with their *c*-axes perpendicular to the stress direction [65,66]. Studies into closely related materials in the tetragonal perovskite $ABO_3$ family have revealed similar domain reorientations [67–71], including simulations on micropillars that reveal an initial domain alignment parallel to the pillar axis, but reorienting perpendicular to the pillar axis under compressive stress [72].

We propose that the reentrant superconductivity in the compressed $SrTiO_3$ micropillar originates from of a 90-degree structural domain reorientation. This interpretation is supported by the discrete, step-like features that were observed in our strontium titanate stress measurements which resemble sizeable domains suddenly reorienting, see Fig. S1 in Supplementary Information. Stress curves in the Ti micropillar that we measures do not exhibit any such discrete steps. In addition, the suppression of the superconducting state is unstable, which is expected if it was dependent on a complex and mobile domain structure. The instability and abruptness of the switch to an enhanced susceptibility also supports the idea of a domain reorientation. At the highest stresses, susceptibility changes to a regular, modest linear increase, which is consistent with the completion of the domain reorientation. Under a 90-degree domain reorientation, the pillar would have an initial zero-stress state that is dominated by a single domain that has the tetragonal *c*-axis parallel to the pillar axis, see Fig. 4b. At low compressive stress, the stress on the pillar in this state will apply compressive strain to the *c*-axis. Thus, the suppression of the superconducting state that we observe in this stress range is the result of compressive strain on the *c*-axis. This is consistent with quantum ferroelectric soft mode models of superconductivity in carrier-doped $SrTiO_3$ because these models predict that *c*-axis compressive strain will stiffen the ferroelectric TO phonon mode, tune the material away from the ferroelectric QCP, and reduce the superconducting coupling constant [4,11]. At high compressive stress, the domains reorient to align perpendicular to the pillar axis, which means that compressive stress on the pillar is now applying tensile strain to the *c*-axis (Poisson's ratio in $SrTiO_3$ is 0.24 [73,74]). The enhancement of the superconducting state in this stress range agrees with quantum ferroelectric soft mode models' prediction that tensile strain on the *c*-axis will soften the ferroelectric TO mode, tune the material towards the QCP and increase the superconducting coupling constant [4,11]. Overall, the susceptibility behavior strongly supports quantum critical ferroelectric models of superconductivity in $SrTiO_3$ when considering domain mechanics.

We chose titanium for our control micropillar because its critical temperature (~0.5 K) is similar to niobium-doped strontium titanate and because low-temperature low-pressure titanium forms a highly isotropic close-packed hexagonal crystal structure [75,76] which is a conventional superconductor [77]. Pressures in our experimental setup were too low to induce the α-ω structural phase transition, which occurs at low temperatures at around 10 GPa [75,76]. Thus, its superconducting state is not expected to have any dependence on stress direction, and domain reorientations under uniaxial stress are also not expected. In our titanium micropillar, there was no observed suppression of the superconducting state at lower pressures, and no enhancement of the superconducting state at higher pressures. Although there were small nonmonotonic features after contact, these features were typically less than 10% of the signal strength just before contact. Some are likely caused by imperfect mechanical contact. The features observed in the Ti pillar are very weak compared to the 100% suppression and 400% enhancement that we observed in the $SrTi_{0.996}Nb_{0.004}O_3$ pillar. These features in the Ti pillar are further distinguished from those in $SrTi_{0.996}Nb_{0.004}O_3$ by their relative weakness at low temperature (high susceptibility) and persistence above $T_c$. Importantly, we did not observe any features in our measurements in the Ti control sample that resemble the extreme suppression and enhancement we observed in $SrTi_{0.996}Nb_{0.004}O_3$, which rules out artifacts such as SQUID loop deformation or SQUID contact heating as causes of the extreme susceptibly changes in $SrTi_{0.996}Nb_{0.004}O_3$.

We interpret the inconsistent history of compression experiments on superconducting $SrTiO_3$ to be the result of a variety of crystal sizes and geometries, which can significantly alter the domain mechanics and change the orientation of the *c*-axis under stress. The majority of past compression experiments on single crystals [37,39] have exhibited a suppressive or mixed response to the superconducting state. We suggest that in bulk samples, superconductivity may be suppressed under compression because domain reorientation is incomplete. In mesoscopic samples such as in this manuscript, proximity to the abundant free surfaces likely enhances domain mobility. Enhanced domain mobility near free surfaces has been directly observed in $PbZr_xTi_{1-x}O_3$ thin films, where domains are strongly pinned by misfit dislocations at substrate interfaces while remaining mobile near free surfaces [78,79]. Relaxation of dislocation arms at surfaces has been a prominent explanation for plastic flow in micropillars [80,81], and it is possible that tetragonal domain walls relax at free surfaces like dislocation arms do during plastic deformation. Another reason that small samples could enable complete domain reorientation is that large domain-pinning defects are statistically unlikely to occur in mesoscopic samples whereas many occur in bulk samples due to their large volume [82]. Finally, we consider the possibility that the initial domain structure makes a significant impact on the evolution of domains under stress, as domain reorientation mechanics have been observed to vary significantly based on initial domain structure [78,83]. The zero-stress state of $SrTiO_3$ is monodomain if the sample is small enough whereas domain structures emerge for larger samples [65], so the difference in the initial domain configuration could be a significant factor in the ability of a sample to undergo a complete domain reorientation. When considering sample size and domain mobility, past experiments on compressed $SrTiO_3$ present a clear picture that bulk samples show a superconducting response indicating low domain mobility. This manuscript shows that compressed mesoscopic samples, conversely, will ultimately enhance the superconducting state because the increased domain mobility will result in an elongation of the *c*-axis.

The results presented in this manuscript challenge the notion that changing the carrier concentration in $SrTiO_3$ can change whether the superconducting state is enhanced or suppressed under compression. Pfeiffer [39] contends that domain switching is not responsible for the variation in the results of their compression experiments on various samples, but supposes instead that it is due to the varying carrier concentrations in the samples. Our sample has a carrier concentration of $6.4\times10^{19}$ $cm^{-3}$, which is the same as the samples of Herrera *et al.* [37] and very similar to Pfeiffer's sample HR33, neither of which exhibited an enhanced superconducting state under uniaxial compression. In fact, it was Pfeiffer's sample Nb20 that exhibited a behavior more consistent with the results of this work. The carrier concentration of sample Nb20 was $2.5\times10^{19}$ $cm^{-3}$ but was reported as having a smaller cross section than sample HR33. The exact dimensions of Pfeiffer's samples are unknown to us and while likely all of them are nominally 'bulk' it is still possible that sample size or another seemingly unimportant detail in sample characteristics could have altered the domain mechanics of the samples. $SrTiO_3$ samples as thick as 0.3 mm have shown significantly different domain mechanics compared to samples a few mm thick [65], so even samples larger than what would typically be considered mesoscopic would be small enough to enable a complete domain reorientation under compression. In micropillars, the typical diameter of 1-10 μm is far below 0.3 mm, so domains are very mobile. Thus, both zero stress and high stress states allow the domains to align in their preferred direction relative to the stress and results are not masked by domain pinning.

Structurally homogenous samples and huge, continuous strains are accessible with the technique presented in this manuscript, which integrates microstructure fabrication with scanning probe microscopy. This technique can be adapted to a wide array of probes and is compatible with many possible samples. PFIB milling is a versatile microstructure fabrication technique that does not require matching a material to a soluble sacrificial layer [84], can be easily performed on any conducting single crystal [85,86], and enables continuous strain, compared to the very discrete strains achievable by thin films [87,88]. We note that this work is concerned with nominally pure crystals as opposed to thin films, which are outside the scope of the present theoretical discussions, as highly strained thin films may exhibit disorder driven effects which remain to be fully explained [88–90]. Overcoming the limitations of past experiments on superconducting $SrTiO_3$ was possible due to a distinct combination of advantages in versatility and accessible strain and demonstrates that the technique we have developed fills an important niche in modern strain engineering of quantum materials.

The large tunable stresses that are accessible with our technique also provide exciting opportunities for future studies with strontium titanate. A careful and thorough mapping of the strain (or other means to induce ferroelectricity) and carrier density superconducting phase diagram has long been eagerly desired in this material. Recent observations [43] and models [21] have described the peak of strontium titanate's superconducting dome within its ferroelectric phase (instead of near the QCP as was theorized previously [4]). These recent developments call for experiments with large tunable strains to access that regime of the phase diagram and map the behavior across the quantum critical point thoroughly. The experiment we present here is an important first step in a successful mapping of this phase diagram, as all relevant regions of the phase diagram should be reachable with a series of micropillars at different carrier concentrations and some modifications to the stiffness of our apparatus to allow for higher stresses. Even though the elastic limit of our strontium titanate micropillar is likely on the order of 10 GPa, in our experiment it achieved a maximum of ~100 MPa of stress. This is about the same pressure which is known to reach the ferroelectric quantum critical point and drive strontium titanate into an ordered ferroelectric state [91]. While we do not rule out the possibility that our micropillar entered a ferroelectric state at the highest stresses we achieved, we have no evidence that it did. We see no significant features in susceptibility at the highest pressures and the pressures at which the suppression (~0-10 MPa) and sharp enhancement (~10-50 MPa) occur in this experiment are much lower than the 100 MPa necessary to induce ferroelectricity. Reports that measured the necessary pressure to induce ferroelectricity utilized samples with high domain mobility and reported domain reorientations [91], and thus there should not be a lower stress required for our micropillars. Such questions invite future studies that access higher stress, access additional carrier concentrations, and develop a measurement of polarization in compressed micropillars at millikelvin temperatures.

In summary, we have performed experiments on superconducting $SrTiO_3$ which integrate scanning SQUID microscopy and micropillars fabricated using a focused ion beam. We measured AC susceptibility which had a nonmonotonic response to compressive stress, wherein the susceptibility is suppressed at low compressive stress but is enhanced at high compressive stress. In addition, we performed the same measurement on a titanium micropillar, which is structurally isotropic, and did not see the same results. Our results on our $SrTi_{0.996}Nb_{0.004}O_3$ micropillar suggest that tetragonal domains in small $SrTiO_3$ samples completely reorient under stress. Considering this domain reorientation, the reentrant behavior we observe in the AC susceptibility supports modern quantum critical superconducting models of doped $SrTiO_3$ because the superconducting state of our micropillar is suppressed when the *c*-axis is compressed but it is enhanced when the *c*-axis is stretched. This connection between superconductivity and domain mechanics also reconciles inconsistent results from previous compression experiments on doped $SrTiO_3$ and highlights the future need for careful consideration of sample size, sample geometry, and domain mechanics for strain-tuned experiments on systems with reduced structural symmetry. The results presented in this manuscript encourage further interest in integrating quantum materials research with mesoscopic samples because the technique we present offers a valuable set of advantages: huge tunable strains, structural homogeneity, and sample versatility. For example, $KTaO_3$ is a similar quantum paraelectric system to $SrTiO_3$ that is superconducting with carrier doping but has a higher unstrained $T_c$ of >1 K and strong spin orbit coupling, which results in a rich array of quantum effects [92]. The technique described in this manuscript poses an exciting opportunity for future experiments on many complex materials such as $KTaO_3$, on which studies of highly tunable strain has thus far been sparse.

**Data availability**

The data supporting the findings of this study are available within the Article, its Supplementary Information, raw data is available from the corresponding author upon a reasonable request.

## Acknowledgements

J.F., J.B., B.P., and I.S. thank Alexander Balatsky for fruitful discussions, and also thank Seok-Woo Lee for introducing us to the micropillar fabrication and mechanical testing methods. The FIB fabrication was performed using the facilities in the UConn/ Thermo Fisher Scientific Center for Advanced Microscopy and Materials Analysis (CAMMA).

**Funding Statement**

J.F., J.B., B.P., and I.S. acknowledge support through the University of Connecticut. I.S. acknowledges support by the National Science Foundation Award No. 2233149. I.S. also acknowledges support from UCONN Quantum Region Partnership Investment Award and Quantum SEED Award. J.F., J.B., B.P., and I.S. gratefully acknowledge the Air Force Research Laboratory, Materials and Manufacturing Directorate (AFRL/RXMS) for support via Contract No. FA8650-21-C5711. Distribution A. Approved for public release: distribution unlimited. (AFRL-2025-1449) Date Approved 03-19-2025.

**Author contributions**

J. F. and I. S. wrote the manuscript. J. F. performed the scanning SQUID measurements. J. B. and J. F. prepared the samples. J. B. assisted with the scanning SQUID experiments. B. P. calibrated the stress response. I. S. conceived and supervised the project.

**Competing interests**

The authors declare no competing interests.

Supplementary Information

for

**Reentrant Superconductivity in Strontium Titanate Micropillars Under Elastic Stress**

**Jacob Franklin[1, ♦], Joshua Bedard[1, ♦], Brenna Petrelli[1] and Ilya Sochnikov[1,2,3, ♣]**

[1]*Department of Physics, University of Connecticut, Storrs, CT, USA 06269*

[2]*Institute of Material Science, University of Connecticut, Storrs, CT, USA 06269*

[3]*Department of Material Science & Engineering, University of Connecticut, Storrs, CT, USA 06269*

♦*Equal contributors*

♣*Corresponding author; e-mail: ilya.sochnikov@uconn.edu*

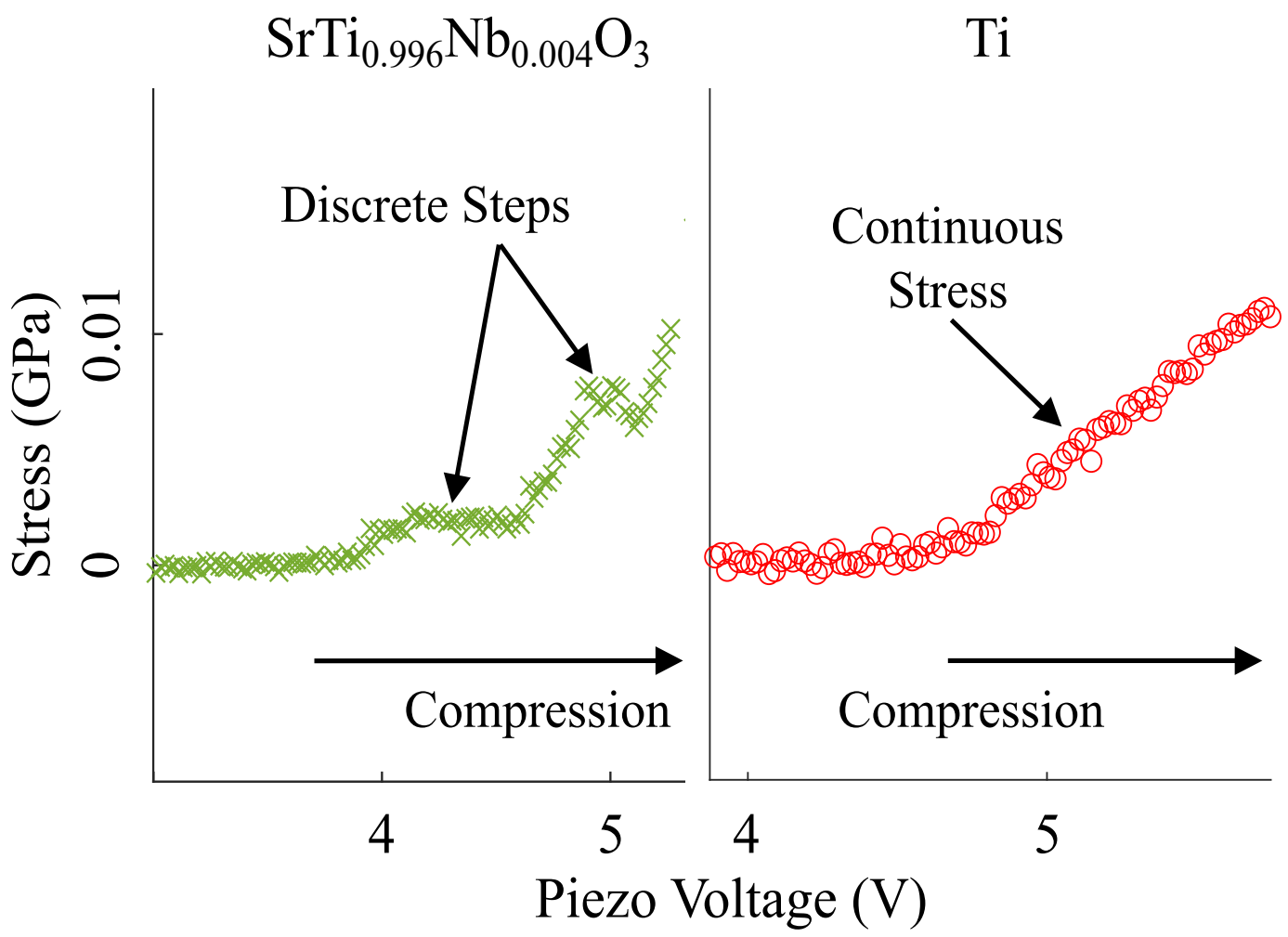

FIG S1. (left) A zoomed-in view of a typical $SrTi_{0.996}Nb_{0.004}O_3$ stress curve from Fig. 2 shows a discrete, step-like behavior that is indicative of sizable tetragonal domains suddenly reorienting. Under lower stress, these domains would be aligned with the tetragonal *c*-axis parallel to the pillar axis but with increasing compressive stress the domains will reorient to align the tetragonal *c*-axis perpendicular to the pillar axis. (right) A zoomed-in view of a typical Ti stress curve from Fig. 3 shows no such discrete steps and exhibits a smooth linear response. Domain reorientations are not expected in Ti under low stress (<10GPa) due to the highly isotropic nature of titanium's close-packed hexagonal α phase.